\documentclass[
 reprint,
showpacs,preprintnumbers,
 amsmath,amssymb,
 aps,
prc,
]{revtex4-1}

\usepackage{graphicx}% Include figure files
\usepackage{dcolumn}% Align table columns on decimal point
\usepackage{bm}% bold math

\begin{document}

\preprint{APS/123-QED}

\title{The neutron star inner crust: an empirical essay}% Force line breaks with \\
%\thanks{A footnote to the article title}%

\author{Luiz L. Lopes}
\email{llopes@cefetmg.br}
\affiliation{Centro Federal de Educa\c c\~ao  Tecnol\'ogica de
  Minas Gerais Campus VIII, CEP 37.022-560, Varginha, MG, Brasil}

\date{\today}% It is always \today, today,
             %  but any date may be explicitly specified

\begin{abstract}
In this work I study how the small contribution of the inner crust to the total equation of state (EoS) of a neutron star affects its mass-radius relation, focusing on the canonical mass of 1.4$M_\odot$. I build a polytropic empirical EoS of the kind: $p = K\epsilon^{\gamma} + b$,  in the range of 0.003 fm$^{-3}~<~n~<~$0.08 fm$^{-3}$ and also calculate the speed of sound in this region. We see that different behaviours of the speed of sound can affect the radius of the canonical star by more than 1.1 km. This result can help us understand extreme results as GW170817,
where some studies indicate that the  radius of the canonical star cannot exceed 11.9 km.
\end{abstract}

% PACS, the Physics and Astronomy
                             % Classification Scheme.

\maketitle

\section{Introduction \label{sec1}}

The physics of neutron stars is an old subject. It can be traced back from the work of the Soviet physicist  Landau~\cite{Landau} in early 1930s, through the pioneering work of Oppenheimer and Volkoff about massive neutron cores in fully relativistic formalism in 1939~\cite{TOV}. The discovery of 
radio pulsars by  Bell and  Hewish~\cite{Bell} in 1967 
made clear that the nuclear forces became repulsive at low distances due to the  pulsar masses.
Nowadays we believe that the EoS of nuclear matter is very stiff at high densities to explain  the recent  discovery of the hyper massive  MSP J0740+6620, whose mass range lies at $2.14^{+0.10}_{-0.09}~M_{\odot}$ with 68$\%$ credibility interval and $2.14^{+0.20}_{-0.18}~M_{\odot}$ with 95$\%$  credibility interval~\cite{Cromartie}, as well the  PSR J0348+0432 whose  mass lies between $2.01 \pm 0.04~M_{\odot}$~\cite{Antoniadis}.

On the other hand, quantitative results about neutron stars radii only significantly evolve in the last decade. The radius of canonical 1.4$M_\odot$ in the past was estimated around 17 km~\cite{17km}. Until today even more conservative results have pointed out  that the radius of canonical stars cannot surpass 14 km~\cite{Hebeler,malik,Riley,Miller,Fatto}. More radical studies point to a maximum radius close to 13 km~\cite{PRL121,Lattimer2013,Lattimer2014},  and
a very recent study constrains the maximum radius to only 11.9 km~\cite{Capano2020}.

We can divide the neutron star in four distinct parts: the outer crust, the inner crust, the outer core and the inner core. It is well accepted
that the symmetry energy slope at the saturation density - which corresponds to the outer core region - is the main responsible for controlling the neutron stars radii.  Although some studies suggest that this cannot be the whole history~\cite{Lopes2014,Lopes2018}, it is undeniable that the symmetry
energy slope plays more than a significant role~\cite{Lattimer2014,Rafa,Gandolfi,Alam,Tsang,Micaela2017}.

In this work, nevertheless, I explore another region of the neutron star: the inner crust. Instead of build a model for it, I study only its behaviour,
using an empirical parametrization for the EoS: $p(\epsilon) = K \cdot \epsilon^{\gamma} + b$, where we varying  the value of $\gamma$ and determine the value of the constants $K$ and $b$ in order to keep the pressure continuous. Also, in order to gain physical insight, I calculated the speed of sound of the inner crust. I show that although for all $\gamma$ values we always have a monotically increasing EoS, we have very distinct behaviour for the speed of sound as well as different values of the radius of the canonical mass.

\section{The neutron star layers.}

As pointed out earlier, the neutron star can be divided in four distinct regions: outer crust, inner crust, outer core and inner core. 

\subsection{Outer crust}

The outer crust is the region considered in the range $10^{-14}$ fm$^{-3}~\lesssim~n~\lesssim~10^{-4}$ fm$^{-3}$, where the ground state of nuclear matter is such that all neutrons
are bound in nuclei, and that it forms a perfect crystal with a single nuclear species, (number of neutrons N, number of protons Z), at lattice sites. The formulation of this model 
is known today as BPS model~\cite{BPS}. For densities up to $10^{-9}$ fm$^{-3}$ the ground state is a body-centered-cubic (bcc) crystal lattice of $^{56}$Fe with negligible - but increasing -
pressure. For higher densities, the matter is a plasma of nuclei and electrons which form a
nearly uniform Fermi gas and the degenerescence pressure of electrons, as well as the lattice pressure - due to the Coulomb interactions - become relevant. For densities above $10^{-7}$ fm$^{-3}$ the  $^{56}$Fe is no longer the ground state of the matter but the  $^{62}$Ni is. 
Therefore, a sequence of increasingly neutron rich nuclei follows,  and ends up in the
$^{78}$Ni for density around $10^{-4}$ fm$^{-3}$~\cite{HaenselBook}.

The ground state composition at a given density corresponds to the absolute
minimum of the baryon chemical potential in the  N-Z plane.
Typically, there is only one well distinguished minimum. A well pronounced
second minimum appears only close to the transition density between two
nucleus species. With increasing density it becomes a new absolute minimum.
(For a complete relation between a given density and the corresponding ground state nucleus,
as well as a longer discussion about the outer  crust see ref.~\cite{HaenselBook}). In this work I use the BPS EoS to the outer crust.

\subsection{Inner crust}

The inner crust is the region comprehending around  $10^{-4}$fm$^{-3}~\lesssim~n~\lesssim~10^{-1}$fm$^{-3}$. Here, very neutron rich nuclei are immersed in a gas of dripped neutrons~\cite{HaenselBook}.

In general, calculations of the structure, composition, and equation of state
of the inner crust can be divided into three groups, alongside many subgroups within different
parametrizations, techniques and approximations: Full quantum mechanical treatment can be carried out within the Hartree-Fock (HF) approximation with an effective nucleon-nucleon interaction as done in ref.~\cite{NV}.
Further approximation of the many-body wave function can be done using semi-classical Extended Thomas-Fermi (ETF) approximation. Basic quantities within the ETF are neutron and proton densities and their spatial gradients, as done in ref.~\cite{ETF}. Finally, investigations belonging to the third group use Compressible Liquid Drop Model (CLDM) parametrization for the description of nuclei, with parameters derived within a microscopic nuclear many-body theory as done
in ref.~\cite{BBP} (called BBP EoS).

Besides the uncertain discussed above, close to the inner crust edge the competition between
attractive nuclear force and repulsive Coulomb interaction can turns the nuclear matter into a 
frustrated system, i.e, the system presents more than one low-energy configuration, which can cause
the onset of unusual  nuclear shapes with different geometries. This is called nuclear pasta phase~\cite{Pasta1983}. As pointed out in ref.~\cite{Lorenz}, the 
presence and extension of the pasta phase is strongly model dependent.
While  ref.~\cite{Douchin} stated that there is no pasta phase, ref.~\cite{Ava2008,Ava2009} shows that pasta phase is present for densities from 
0.006 fm$^{-3}$ to 0.1 fm$^{-3}$.

As can be seen from the discussion above, the inner crust up to date still present significant ambiguities. Therefore, instead of analyze different
models alongside several parametrizations, with and without pasta phase, I construct here an
empirical study of the inner crust for  0.003 fm$^{-3}~<~n<~$0.08 fm$^{-3}$ within a polytropic  EoS:

\begin{equation}
 p(\epsilon) = K.\epsilon^\gamma + b , \label{e1}
\end{equation}
where $p$ is the pressure and $\epsilon$ is the energy density.

Now, it is very important to emphasize  that the pressure needs be kept continuum in the outer crust-inner crust transition, as well as in the inner crust-core transition. To accomplish this task,  for a given value of $\gamma$ we need to reproduce $p = 1.509 \times 10^{-5}$ fm$^{-4}$ for
$\epsilon =  1.402 \times 10^{-2}$ fm$^{-4}$, which is predicted by the BBP EoS at $n = $ 0.0029 fm$^{-3}$~\cite{BBP};
and $p = 2.409 \times 10^{-3}$ fm$^{-4}$ for $\epsilon = 3.838 \times 10^{-1}$ fm$^{-4}$ at $n = $ 0.083
fm$^{-3}$, which is predicted by the Quantum Hadrodynamics (QHD) model NL$\rho$~\cite{Liu} (discussed below). This assures a continuous  pressure.

It is also worth to pointing out that this is not the only possible approach. For instance, ref.~\cite{Hebeler} uses two different polytropic EoS to kept the pressure continuous.

As the NL$\rho$ model predicts a maximum neutron star mass at $n = $ 1.07 fm$^{-3}$, we see that the
analyzed region is just a small fraction of the total EoS. The parameters used in this work are displayed
in Tab.~\ref{T1}

\begin{table}[ht]
\begin{tabular}{|c|c|c|}
\hline
 $\gamma$ & $K$ & $b$    \\
 \hline 

  1/3         & 4.930 $\times~10^{-3}$ & -1.173 $\times~10^{-3}$    \\
 
 2/3          & 5.094 $\times~10^{-3}$ & -2.811 $\times~10^{-4}$    \\

 7/3          & 2.238 $\times~10^{-2}$ & 1.403 $\times~10^{-5}$     \\

 4          & 1.103 $\times~10^{-1}$ & 1.509 $\times~10^{-5}$     \\

 6          & 7.490 $\times~10^{-1}$ & 1.509 $\times~10^{-5}$     \\
\hline

\end{tabular}
 \caption{Parameters of the empirical model of the inner crust in the region  0.003 fm$^{-3}~<~n<~$0.08 fm$^{-3}$.}\label{T1}
 \end{table}

 \subsection{The outer and inner core}
 
 If the density is high enough (around 0.06 - 0.1 fm$^{-3}$) the surface and Coulomb contributions can be
 ignored and the matter can be approximated by an infinite and uniform plasma of interacting protons, neutrons and free electrons (and muons if the electron Fermi energy is high enough) in chemical equilibrium. This is the outer core. Therein  a very special point exists: the nuclear saturation density $n_0$ (0.148 - 0.170 fm$^{-3}$). From this point, the nuclear forces become  repulsive instead of attractive. Any reliable model for the outer core needs to
 predict at least six well known properties of symmetric nuclear matter at the saturation point:
 the saturation density itself ($n_0$), the effective nucleon mass ($M^{*}/M$), the compressibility $(K)$, the symmetry energy ($S_0$),  the binding energy per baryon ($B/A$) and the symmetry energy slope  ($L$)~\cite{Micaela2017,Dutra2014}. Alongside these six physical quantities, ref.~\cite{Daniel} also constrains the EoS in the range 2.0 $~<n/n_0~<$ 4.5.

 To fulfill these constraints I use an extended version of the QHD~\cite{Serot} whose Lagrangian reads:

\begin{widetext}
\begin{eqnarray}
\mathcal{L}_{QHD} = \sum_b \bar{\psi}_b \bigg [\gamma^\mu(i\partial_\mu  - g_{b,\omega}\omega_\mu  - g_{b,\rho} \frac{1}{2}\vec{\tau} \cdot \vec{\rho}_\mu)
- (m_b - g_{b,\sigma}\sigma ) \bigg ]\psi_b     + \frac{1}{2} m_v^2 \omega_\mu \omega^\mu 
   \nonumber \\ + \frac{1}{2} m_\rho^2 \vec{\rho}_\mu \cdot \vec{\rho}^{ \; \mu}   + \frac{1}{2}(\partial_\mu \sigma \partial^\mu \sigma - m_s^2\sigma^2)  
    - U(\sigma)  - \frac{1}{4}\Omega^{\mu \nu}\Omega_{\mu \nu} -  \frac{1}{4}\bf{P}^{\mu \nu} \cdot \bf{P}_{\mu \nu}  , \label{EL1} 
\end{eqnarray}
\end{widetext}
in natural units. 
 $\psi_b$  are the nucleon  Dirac fields.
 The $\sigma$, $\omega_\mu$ and $\vec{\rho}_\mu$ are the mesonic fields.  The $g's$
 are the Yukawa coupling constants that simulate the strong interaction,
 $m_b$ is the mass of the baryon $b$, $m_s$, $m_v$,  and $m_\rho$ are
 the masses of the $\sigma$, $\omega$, and $\rho$ mesons respectively.

 The $U(\sigma)$ is the self-interaction term introduced in ref.~\cite{Boguta} to reproduce some of the saturation properties of the nuclear matter and is given by:
 \begin{equation}
U(\sigma) =  \frac{1}{3!}\kappa \sigma^3 + \frac{1}{4!}\lambda \sigma^{4} \label{EL2} .
\end{equation}

As neutron stars  are stable macroscopic objects, we need to describe a
neutral, chemically stable  matter and hence, leptons are added as free Fermi gases.
In Tab. \ref{T2}, I display the parameters of the NL$\rho$ model
as well as the prediction of the physical quantities and their inferred values from
phenomenology~\cite{Dutra2014,Micaela2017}. Moreover, as pointed out in ref.~\cite{Rafa,Dutra2014},
the EoS constraint from ref.~\cite{Daniel} is also fully satisfied. It also predicts a maximum mass larger than two solar mass, in agreement with ref.~\cite{Cromartie,Antoniadis}.

\begin{figure*}[ht]
\begin{tabular}{cc}
\includegraphics[width=5.6cm,height=7.0cm,angle=270]{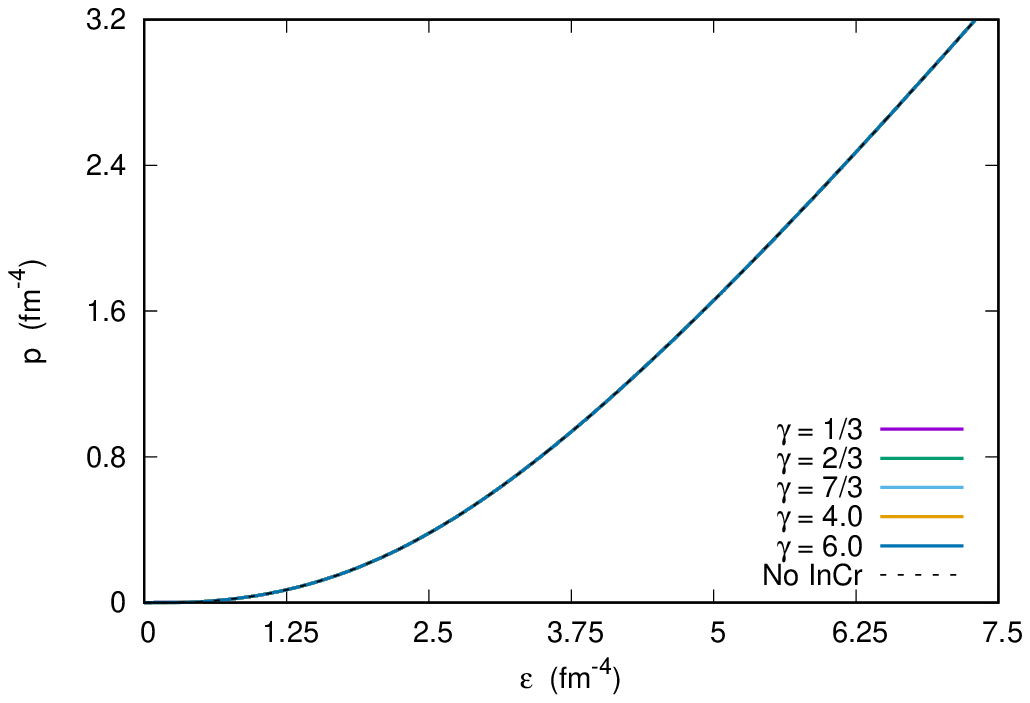} &
\includegraphics[width=5.6cm,height=7.0cm,angle=270]{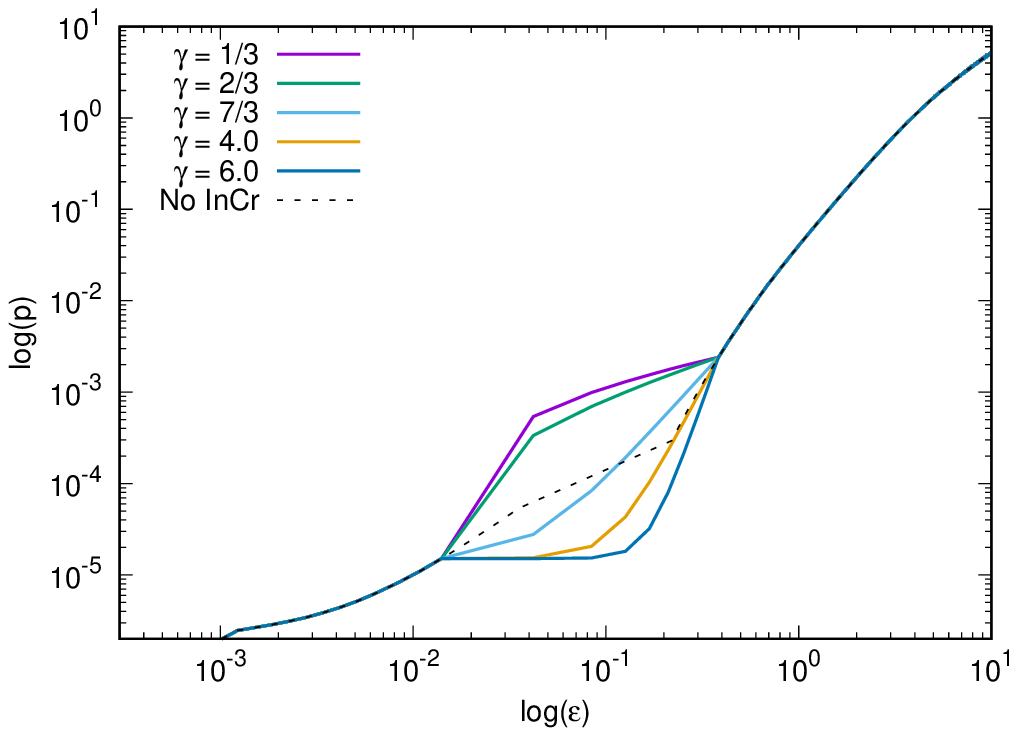} \\
\end{tabular}
\caption{(Colour online) Different EoS for the inner crust in linear (left) and logarithm (right) scales. The differences cannot be perceived in the linear scale. } \label{FL1}
\end{figure*}

\begin{widetext}
\begin{center}
\begin{table}[ht]
\begin{center}
\begin{tabular}{|c|c||c|c|c|}
\hline 
  & Parameters & &  Phenomenology  & NL$\rho$  \\
 \hline
 $(g_{N\sigma}/m_s)^2$ & 10.330 $fm^2$ &$n_0$ ($fm^{-3}$) & 0.148 - 0.170 & 0.16 \\
 \hline
  $(g_{N\omega}/m_v)^2$ & 5.421  $fm^2$ & $M^{*}/M$ & 0.7 - 0.8 & 0.75  \\
  \hline
  $(g_{N\rho}/m_\rho)^2$ & 3.830  $fm^2$ & $K$ (MeV)& 200 - 260 &  240  \\
 \hline
$\kappa/M_N$ & 0.01387 & $S_0$ (MeV) & 30 - 34 &  30.49   \\
\hline
$\lambda$ &  -0.0288 & $B/A$ (MeV) & 15.7 - 16.5 & 16.0 \\
\hline 
$M_N$ &  939 MeV & $L$ (MeV) & 36 - 86.8 & 84.9 \\ 
\hline
\end{tabular}
 
\caption{NL$\rho$ model  parameters and predictions~\cite{Liu} with the physical quantities inferred from experiments
  \cite{Dutra2014, Micaela2017}. } 
\label{T2}
\end{center}
\end{table}
\end{center}
\end{widetext}

The region with $n~>2~n_0$ is called inner core. At such densities new and exotic degrees
of freedom can be present. The most common non-nucleonic degrees of freedom studied
in the literature are the hyperons~\cite{HaenselBook,Glen,Ellis,Rafa2005,Weiss2,Lopes2013,Lopes2012}.
Another possibility is the onset of $\Delta's$ resonances and boson condensation~\cite{Glen,Dexhemier}. Even more exotic settings consider that the inner core undergoes
a hadron-quark phase transition. In this case we have a hybrid star, with a quark gluon plasma (QGP)
at the inner core surrounded by baryonic matter~\cite{HaenselBook,Glen,Annala,Sandoval,Lopes2020,Lopes2020b}. In this work I consider only nucleonic degrees of freedom, therefore the outer and the inner core can be considered as the same
layer.

Furthermore, the details of the  construction of a beta-stable  EoS from the QHD at mean field approximation used in this work can easily
be found in the literature~\cite{HaenselBook,Serot,Glen,Lopes2012,Lopes2020b}.

\subsection{Results}

I construct the total EoS for the neutron stars as follows: We use the BPS~\cite{BPS}
EoS for the outer crust and the BBP~\cite{BBP} EoS for the inner crust for densities
up to 0.003 fm$^{-3}$. The EoS for the core is given by QHD calculations with NL$\rho$
parametrization~\cite{Liu} starting at 0.08 $fm^{-3}$. For this model, this corresponds
exactly to half of the saturation density. The small region in the inner crust in the range 0.003 fm$^{-3}$ $<~n~<$ 0.08 fm$^{-3}$ is parametrizated by the eq.~\ref{e1}. I also compare the results without this parametrization. In this case the BBP EoS is used for densities up to  0.035 fm$^{-3}$  and linked directly to the QHD EoS. This is the standard approach~\cite{HaenselBook,Glen}, here I  called it ``BPS+BBP''. 

I displayed in Fig.~\ref{FL1} the EoS for all values of $\gamma$. As pointed out earlier,
the analysed  region is very small, therefore the differences between the different values
of $\gamma$ cannot be perceived with the traditional linear plot. The differences become
clear if we use use logarithm scale. As can be seen, the lower the value of $\gamma$, the stiffer the EoS for the inner crust. When we link the BBP to the QHD directly, (BPS+BBP) we see that that this model is similar to $\gamma$ = 7/3 up to energy density of 0.01 fm$^{-4}$; afterwards it becomes closer to $\gamma$ = 4.

\begin{figure}[htb] 
\begin{centering}
 \includegraphics[angle=270,
width=0.44\textwidth]{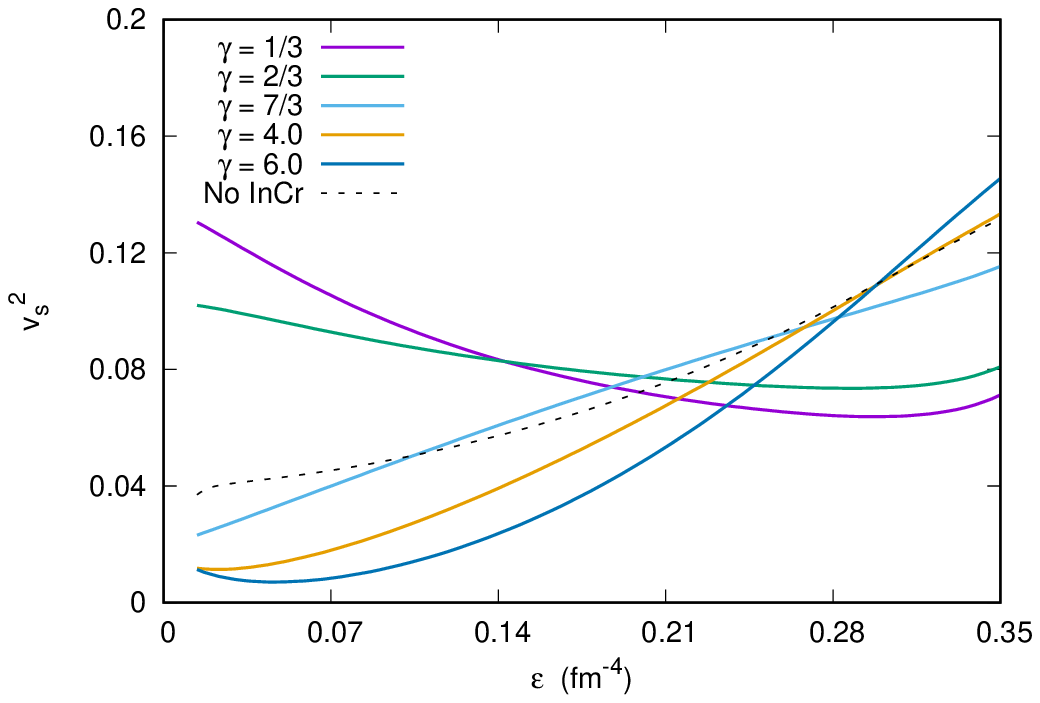}
\caption{(Colour online) Square of the speed of sound in the parameterizated inner crust for different values of $\gamma$. } \label{FL2}
\end{centering}
\end{figure}

As for all values of $\gamma$ we always have a monotically increasing EoS, we can gain additional physical insight studying the behaviour of the speed of sound of the inner crust.
The square of the speed of sound is defined as:
\begin{equation}
 v_s^2 =  \bigg | \frac{\partial p}{\partial \epsilon } \bigg |. \label{e4} 
\end{equation}

Besides the effect of different speeds of sounds of the crust on the mass-radius relations,
the speed of sound also provides us information about shear viscosity~\cite{Epstein}, tidal
deformation~\cite{Pegios} and even gravitational waves signatures~\cite{Miguel}. At high
densities the speed of sound is also relevant in the study of hadron-quark phase transition~\cite{Annala}. The results are showed in Fig.~\ref{FL2}.

We see that the differences in the speed of sound are bigger than the EoS itself.
For low values of $\gamma$ we have a decrease of the speed of sound with the density. As for higher values, we have a quickly increasing speed of sound. In the case in which  we have the BBP EoS linked to the QHD (BPS+BBP) we see the same behaviour of the EoS. At low energy density the 
speed of sound is close to $\gamma$ = 7/3 and therefore closer to $\gamma$ = 4.

Also, as pointed out in ref.~\cite{AIP}, we expect a first order phase transition at the Inner Crust-core surface. Therefore, unlike the pressure, the speed of sound can present discontinuity.

Once I discussed the behaviour of the EoS and the speed of sound of the inner crust I
finish the task by solving the TOV equations~\cite{TOV} to see how these different 
behaviours affects the mass-radius relation to the maximum mass and the canonical 1.4$M_\odot$.  Finally, with the mass-radius relation, I also calculate the gravitational surface redshift, defined as~\cite{Hebeler}:

\begin{equation}
 z = \bigg ( 1 - \frac{2GM}{Rc^2} \bigg )^{-1/2} - 1 . \label{red}
\end{equation}

\begin{figure}[htb] 
\begin{centering}
 \includegraphics[angle=270,
width=0.44\textwidth]{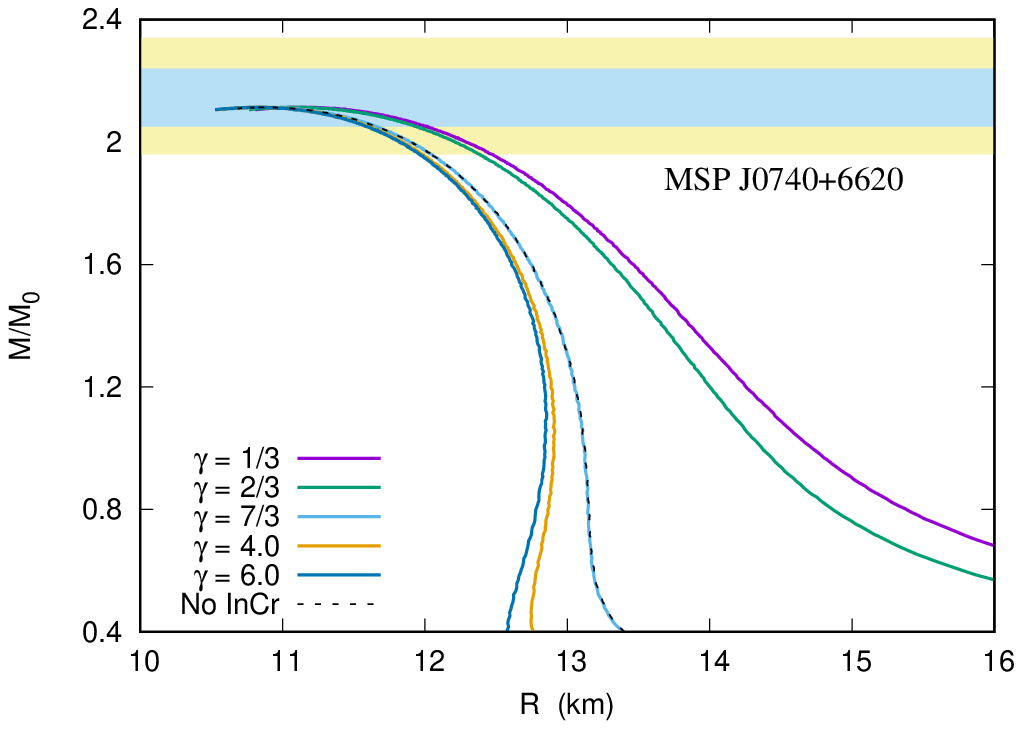}
\caption{(Colour online) Mass-radius relations for different values of $\gamma$.The light blue (yellow)  region corresponds to the credibility interval of 68$\%$ (95$\%$).} \label{FL3}
\end{centering}
\end{figure}

The results are plotted in fig.~\ref{FL3} and summarized in Tab.~\ref{T3}. As can be seen, my results point to the fact  that while the inner crust does not affect the maximum mass  (indeed, the central density for all maximum mass stars is 1.07 fm$^{-3}$), it can plays 
a major role in the radii, specially in the canonical 1.4$M_\odot$. For lower values of 
$\gamma$, we found  big radii for the canonical masses, reaching 13.87 km. For $\gamma$ = 6 we found a radius of 12.75 for the 1.4 solar mass star, a difference of 1.1 km. For the "BPS+BBP'' we see that the results are numerically equivalent to $\gamma$ = 7/3.  I also
checked the inner crust effects with different parametrizations (as GM1~\cite{Glen2}) and found the same qualitative results.

\begin{table}[ht]
\begin{tabular}{|c|c|c||c|c|c|}
\hline
 $\gamma$  & $M/M_\odot$ & R (km) & $z_{1.4}$  & R$_{(1.4)}$ & $n_{c_{(1.4)}}$    \\
\hline
 1/3       & 2.11 & 11.13   & 0.195  & 13.87 & 0.42   \\
\hline
 2/3        & 2.11 & 11.08  & 0.199 & 13.66 & 0.43   \\
 \hline
7/3         & 2.11 & 10.86  & 0.214 & 12.93 & 0.43   \\
\hline
4          & 2.11 & 10.82  & 0.216 & 12.81 & 0.43  \\
\hline
6         & 2.11 &  10.81 & 0.218 & 12.75 & 0.43   \\
\hline
BPS+BBP    &   2.11   & 10.88 & 0.214  & 12.93 & 0.43   \\
\hline
\end{tabular}
 \caption{Some neutron star properties for the  models discussed in the text.}\label{T3}
 \end{table}
 
  On the other hand, the gravitational redshift for the canonical mass, $z_{1.4}$, varies from 0.195 to 0.218, a difference about of 10$\%$. These values for $z$ are in agreement of observational measures coming form the Chandra X-Ray observatory. For instance, ref.~\cite{Sanwal} points out that the redshift $z$ of the  1E 1207.4-5209 pulsar lies between 0.12 to 0.23, although its mass and radius values are unknown.

As I fix the EoS for the neutron star core and varying only the EoS of the crust, we can compare my results with studies that make exactly the opposite: fix the EoS crust while varying
the EoS of the core as done in ref.~\cite{Rafa,Lopes2014}. Both studies vary the 
symmetry energy slope $L$ while keeping all the other physical quantities
unchanged and also obtain similar maximum masses. For instance, ref.~\cite{Rafa}, using non-linear $\omega$-$\rho$ coupling found that the radius of  canonical mass
can vary from 12.55 to 13.41 km, a difference of around 0.9 km (Table V of ref.~\cite{Rafa}). On other hand
ref.~\cite{Lopes2014} use the additional scalar-isovector $\delta$ meson and the same
NL$\rho$ model as I use here;  it found that the radius of the canonical star 
varying from 12.97 to 14.10, a difference of 1.1 km (Table 16 of ref.~\cite{Lopes2014}).
We see that the differences in the radii varying the inner crust EoS are very close
to the works that vary the core Eos. This is quite impressive, as the inner crust 
of a canonical star is less than 20$\%$ of the total EoS.

\subsection{Conclusions}

In this work I build an  empirical EoS for the inner crust and study how different
behaviours affect the canonical mass neutron stars.  I found that
different behaviours in the speed of sound produce very different radii.

 The strong constraint imposed by the  GW170817, where the radius of the  canonical mass cannot exceed 11.9 km~\cite{Capano2020} indicates that not only the homogeneous core EoS, but also the inhomogeneous  crust EoS are needed to be known with a high degree of precision. Ref.~\cite{Rafa,Lopes2014} indicates that a small symmetry energy slope ($L$) value can reduce the radius of the canonical star of about 1.1 km. On the other hand,  here I show that  a inner crust EoS  within a quickly increasing speed of sound, also reduces the radius og the same order, an impressive result as the core EoS correspond to more than 80$\%$ of the total EoS.

It is also worth to point that ref.~\cite{Patra} study the effects of the inner crust using  Thomas-Fermi approximation. The authors vary the parametrizations of the inner crust EoS, but  fixed a single  parametrization for the core EoS. They show that the variations of the inner crust can affect the radius of the  canonical star up to 0.7 km.
This corroborate the results pointed out in this work,  that the inner crust effects are non-negligible. Nevertheless, no parameterization was able to reproduce the 11.9 km radius constraint of the  GW170817~\cite{Capano2020}.

\end{document}